\documentclass{aa}

\usepackage{txfonts}
\usepackage{graphicx,amsmath,amssymb}
\usepackage{color}
\usepackage[switch]{lineno} 
\usepackage{natbib}

\bibpunct{(}{)}{;}{a}{}{,}

\usepackage[colorlinks=true,citecolor=blue,urlcolor=blue,linkcolor=blue]{hyperref}

\begin{document}

\title{
Multigroup radiation hydrodynamics with flux-limited diffusion and adaptive mesh refinement
}

\author{
  M.~Gonz\'alez\inst{1}, N. Vaytet\inst{2}, B. Commer\c{c}on\inst{2}, J.~Masson\inst{3,2}
}

\authorrunning{M. Gonz\'alez et al.}
\titlerunning{Multigroup radiation hydrodynamics with adaptive mesh refinement}

\institute{
  Univ Paris Diderot, Sorbonne Paris Cit\'e, AIM, UMR7158, CEA, CNRS, F-91191
    Gif-sur-Yvette, France \and
  \'Ecole Normale Sup\'erieure de Lyon, CRAL, UMR CNRS 5574, Universit\'e de Lyon, 46 All\'ee d'Italie, 69364 Lyon Cedex 07, France \and
  School of Physics, University of Exeter, Exeter EX4 4QL, UK
}

\date{Received XX / Accepted XX}

\abstract
{Radiative transfer plays a key role in the star formation process. Due to a high computational cost, radiation-hydrodynamics simulations performed up to now have mainly been carried out in the grey approximation. In recent years, multi-frequency radiation-hydrodynamics models have started to emerge, in an attempt to better account for the large variations of opacities as a function of frequency.
}
{We wish to develop an efficient multigroup algorithm for the adaptive mesh refinement code \texttt{RAMSES} which is suited to heavy proto-stellar collapse calculations.
}
{Due to prohibitive timestep constraints of an explicit radiative transfer method, we constructed a time-implicit solver based on a stabilised bi-conjugate gradient algorithm, and implemented it in \texttt{RAMSES} under the flux-limited diffusion approximation.
}
{We present a series of tests which demonstrate the high performance of our scheme in dealing with frequency-dependent radiation-hydrodynamic flows. We also present a preliminary simulation of a three-dimensional proto-stellar collapse using 20 frequency groups. Differences between grey and multigroup results are briefly discussed, and the large amount of information this new method brings us is also illustrated.
}
{We have implemented a multigroup flux-limited diffusion algorithm in the \texttt{RAMSES} code. The method performed well against standard radiation-hydrodynamics tests, and was also shown to be ripe
for exploitation in the computational star formation context.
}


\maketitle

\section{Introduction}

Numerical studies of star formation are very demanding, as many physical mechanisms need to be taken into account (hydrodynamics, gravity, magnetic fields, radiative transfer, chemistry, etc). 
Models are rapidly increasing in complexity, providing ever-more realistic interpretations of todays highly advanced observations of protostellar systems, such as the recent groundbreaking images taken by the ALMA
interferometer\footnote{\scriptsize{\url{http://www.almaobservatory.org/en/press-room/press-releases/771-revolutionary-alma-image-reveals-planetary-genesis}}}.
Radiative transfer plays a key role in star formation; it acts as a conduit to remove compressional heating during the initial stages of cloud collapse, enabling an isothermal contraction 
\citep{Larson1969,Masunaga1998}, and it also inhibits cloud fragmentation in large-scale simulations \citep[see][for instance]{Bate2012}.
State-of-the-art simulations thus require the solutions to the full radiation magneto-hydrodynamics (RMHD) system of equations, and three-dimensional simulations have only just recently become 
possible with modern computers \citep[see][for example]{Commercon2011c,Tomida2013}.
In particular, including frequency dependent radiative transfer is essential to properly take into account 
the strong variations of the interstellar gas and dust opacities as a function of frequency \citep[see for example][]{Ossenkopf1994,Li2001,Draine2003,Semenov2003,Ferguson2005}.
Three-dimensional calculations including full frequency-dependent radiative transfer are still out of reach of current computer architectures.

In order to overcome this difficulty, much effort has been spent in recent years developing mathematically less complicated, yet accurate approximations to the equations of radiative transfer. 
Such representations include diffusion approximations, the M$_{1}$ model, 
short and long characteristics methods,
and Variable Eddington Tensor descriptions.
One of the simplest and most widely used is the
flux-limited diffusion (FLD) approximation \citep{Levermore1981}, and has been applied to many areas of physics and astrophysics.

These methods tend to drastically reduce computational cost, but still they are often integrated over all frequencies (also known as the `grey' approximation) as the multi-frequency formalism remains too expensive. Only in recent 
years have multigroup methods (whereby the frequency-dependent quantities are binned into a finite number of groups and averaged over the group extents) appeared in numerical codes
\citep[][to mention a few]{Shestakov2008,Vanderholst2011,Vaytet2011,Davis2012,Zhang2013,Vaytet2013b}, as a first step towards accounting for the frequency dependence of gas and dust quantities in calculations.
Multigroup methods are very suited to astrophysics as they allow adaptive widths of groups, thus enabling to use a small number of groups, concentrating frequency resolution where it is
needed (i.e. mostly where opacity gradients are strong). Group boundaries are usually chosen once at the start of the simulations, but more complex schemes have also been written with moving 
adaptive borders \citep{Williams2005}, as absorption and emission coefficients generally vary with the material temperature and density.

\citet{Commercon2011b} implemented frequency-integrated
FLD in the adaptive mesh refinement (AMR) code \texttt{RAMSES} \citep{Teyssier2002,Fromang2006}, and devised an adaptive time-stepping scheme in a follow-up paper \citep{Commercon2014}. In
this third paper, we extend the method to a multigroup formalism. Even though multigroup effects were found to only have a small impact on one-dimensional (1D) simulations of star formation \citep{Vaytet2012,Vaytet2013a},
they are expected to be enhanced in 3D where the optical thickness can markedly vary along different lines of sight \citep[see][]{Kuiper2011}. Using a frequency-dependent method is also the only
way to correctly model ultra-violet radiation from stars being absorbed by surrounding dust and re-emitted in the infrared. Finally, multigroup formalisms, compared to grey methods, are known 
to significantly affect the structures of radiative shocks \citep{Vaytet2013b}, alter energy transport in stellar atmospheres \citep{Chiavassa2011}, as well as being essential to neutrino 
transport in core collapse supernovae explosions \citep[see][for instance]{Mezzacappa1998}.

We first present the numerical method we have used, followed by a series of tests against analytical solutions, and we end with an application to the collapse of a 
gravitationally unstable cloud, comparing grey and multigroup results.

\section{Numerical method} \label{sec:num}

The FLD multigroup radiation hydrodynamics equations
in the frame comoving with the fluid are
\begin{equation}
\begin{array}{@{}l@{~~}c@{~~}l@{}}
 \partial_t \rho + \nabla \cdot [ \rho {\bf u} ] & = & 0 \\
 \partial_t (\rho {\bf u}) + \nabla \cdot [ \rho {\bf u} \otimes {\bf u} + P \mathbb{I} ] & = & -\displaystyle \sum_{g=1}^{Ng}{\lambda_g \nabla E_g} \\
 \partial_t E_{\mathrm{T}} + \nabla \cdot [ {\bf u} (E_{\mathrm{T}} + P) ] & = & \displaystyle \sum_{g=1}^{Ng}{\left[- \mathbb{P}_g : \nabla {\bf u} - \lambda_g {\bf u} \cdot \nabla E_g \right.} \\
 && \left. + \nabla \cdot \left(\displaystyle \frac{c\lambda_g}{\rho \kappa_{\mathrm{R}g}} \nabla E_g\right) \right]\\
 \partial_t E_g + \nabla \cdot [ {\bf u} E_g ] & = & - \mathbb{P}_g : \nabla {\bf u} + \nabla \cdot \left(\displaystyle \frac{c\lambda_g}{\rho \kappa_{\mathrm{R}g}} \nabla E_g\right) \\
 && + \kappa_{\mathrm{P}g} \rho c \left(\Theta_g(T)-E_g\right) \\
 && + \nabla {\bf u} : \displaystyle \int_{\nu_{g-1/2}}^{\nu_{g+1/2}}{\partial_\nu(\nu \mathbb{P}_\nu) d\nu}
\end{array}
\label{eq:rhd}
\end{equation}
where $c$ is the speed of light, $\rho$, $\bf{u}$, $P$, and $T$ are the gas density, velocity, pressure, and temperature, respectively. $E_{\mathrm{T}}$ is the total energy $E_{\mathrm{T}}=\rho \epsilon + 1/2 \rho u^2 + \sum_{g=1}^{Ng}{E_g}$
($\epsilon$ is the internal specific energy), and $\mathbb{I}$ is the identity matrix. We also define
\begin{equation}\label{eq:groupvar}
X_{g} = \int_{\nu_{g-1/2}}^{\nu_{g+1/2}} X_{\nu} d\nu
\end{equation}
where $X = E$, $\mathbb{P}$ which represent the radiative energy and pressure inside each group $g$ which holds frequencies between $\nu_{g-1/2}$ and $\nu_{g+1/2}$.
$N_{g}$ is the total number of groups, $\Theta_{g}(T)$ is the energy of the photons having a Planck distribution at temperature $T$ inside a given group.
The coefficients $\lambda_g$, $\kappa_{\mathrm{P}g}$ and $\kappa_{\mathrm{R}g}$ are, respectively, the flux-limiter, the Planck and the Rosseland means of the spectral opacity $\kappa_{\nu}$ inside a given group.

We employ a commonly used operator splitting scheme whereby the equations of hydrodynamics are first solved explicitly using the second order Godunov method of \texttt{RAMSES} including the 
radiative terms involving $\nabla \cdot \mathbf{u}$ and $\nabla E_{g}$, while the evolution of the radiative energy density and its coupling to the gas internal energy is solved implicitly
\citep[for more details on the different equations which are solved explicitly and implicitly, we refer the reader to the exhaustive description in][]{Commercon2011b}.

To discretize the equations solved in the implicit step, we linearize the source term following
\begin{equation}\label{eq:linsource}
\Theta_g(T^{n+1})=\Theta_g(T^n)+\Theta'_g(T^n)(T^{n+1}-T^{n}) ~,
\end{equation}
where the prime denotes the derivative with respect to temperature. This then enables us to write a set of discretised equations, expressed here in 1D for simplicity, for the evolution of the gas temperature (where $C_{\mathrm{v}}$ is the gas heat capacity at constant volume)
\begin{equation}\label{eq:discrTemp}
T_i^{n+1}=\frac{{C_{\mathrm{v}}^{n}}_i T_i^n-\sum_g{{\kappa_{\mathrm{P}}}_{g,i}^{n}\rho_i^{n} c \Delta t \left( \Theta_g(T_i^n)-T_i^n \Theta'_g(T_i^n)-E_{g,i}^{n+1}\right)}}{{C_{\mathrm{v}}^{n}}_i+\sum_g{{\kappa_{\mathrm{P}}}_{g,i}^{n}\rho_i^{n} c \Delta t \Theta'_g(T_i^n)}}
\end{equation}
and the radiative energy
\begin{equation}\label{eq:discrEr}
\begin{array}{@{}l@{}}
\displaystyle E_{g,i}^{n+1}\left[ 1+ {\kappa_{\mathrm{P}}}_{g,i}^{n}\rho_i^{n} c \Delta t + \frac{c\Delta t}{V_i} \left( \frac{\lambda_g}{\rho^{n}{\kappa_{\mathrm{R}}}_g^{n}}\frac{S}{\Delta x}\right)_{i-1/2} +  \frac{c\Delta t}{V_i}\left( \frac{\lambda_g}{\rho^{n}{\kappa_{\mathrm{R}}}_g^{n}}\frac{S}{\Delta x}\right)_{i+1/2} \right] \\
\displaystyle - \frac{c\Delta t}{V_i}  \left(\frac{\lambda_g}{\rho^{n}{\kappa_{\mathrm{R}}}_g^{n}}\frac{S}{\Delta x}\right)_{i-1/2} E_{g,i-1}^{n+1} - \frac{c\Delta t}{V_i}  \left(\frac{\lambda_g}{\rho^{n}{\kappa_{\mathrm{R}}}_g^{n}}\frac{S}{\Delta x}\right)_{i+1/2} E_{g,i+1}^{n+1} \\
\displaystyle - {\kappa_{\mathrm{P}}}_{g,i}^{n}\rho_i^{n} c \Delta t \Theta'_g(T_i^n) \sum_{\alpha}{\frac{{\kappa_{\mathrm{P}}}_{\alpha,i}^{n}\rho_i^{n} c \Delta t}{{C_{\mathrm{v}}^{n}}_i+\sum_{\beta}{{\kappa_{\mathrm{P}}}_{\beta,i}^{n}\rho_i^{n} c \Delta t \Theta'_{\beta}(T_i^n)}} E_{\alpha,i}^{n+1}} \\
\displaystyle ~~~~~~~~~~~~~~~~~~~~~~~~~~~~~~ = E_{g,i}^n + {\kappa_{\mathrm{P}}}_{g,i}^{n}\rho_i^{n} c \Delta t \left( \Theta_g(T_i^n)-T_i^n \Theta'_g(T_i^n) \right) \\
\displaystyle + {\kappa_{\mathrm{P}}}_{g,i}^{n}\rho_i^{n} c \Delta t \Theta'_g(T_i^n) \frac{{C_{\mathrm{v}}^{n}}_i T_i^n-\sum_{\alpha}{{\kappa_{\mathrm{P}}}_{\alpha,i}^{n}\rho_i^{n} c \Delta t \left( \Theta_{\alpha}(T_i^n)-T_i^n \Theta'_{\alpha}(T_i^n)\right)}}{{C_{\mathrm{v}}^{n}}_i+\sum_{\alpha}{{\kappa_{\mathrm{P}}}_{\alpha,i}^{n}\rho_i^{n} c \Delta t \Theta'_{\alpha}(T_i^n)}} ~.
\end{array}
\end{equation}
The terms with superscripts $n+1$ refer to the variables evaluated at the end of the timestep $\Delta t$, while superscripts $n$ indicate the state at the beginning of the timestep. Subscripts 
$i$ represent the grid cell, and $i \pm 1/2$ are for cell interfaces. In addition, $V$, $S$, and $\Delta x$ are the cell volume, the interface surface area, and the cell width, respectively.
The subscripts $\alpha$ and $\beta$ denote the frequency groups where the subscript $g$ is already in use.

The implicit step requires the inversion of a large matrix which holds the system of equations (\ref{eq:discrTemp}) and (\ref{eq:discrEr}), and this is performed using a parallel iterative 
method\footnote{The use of a direct inversion method is not suited to the very large systems of equations that need to be solved in heavy 3D simulations.}.
In the case of the grey approach (and in our specific case of Cartesian coordinates), the matrix to invert in the implicit step is symmetric, which allows to use the conjugate gradient method
\citep[for further details, see][]{Commercon2011b,Commercon2014}.
However, in the multigroup case, the interaction between radiative groups adds non-symmetric terms, for which we had to implement a similar but more advanced
stabilised bi-conjugate gradient (BiCGSTAB) algorithm \citep{VanderVorst1992}.

Finally, the term which depends on a derivative with respect to frequency ($\partial_{\nu}$) accounts for energy exchanges between neighbouring groups due to Doppler effects. It is computed
using the method
described in \citet{Vaytet2011}, and treated explicitly as part of the final line in equation (11) in \citet{Commercon2011b} (note that we now have one such line per frequency group).

\section{Method validation} \label{sec:tests}

We present in this section the numerical tests performed to assess the accuracy of our method.

\subsection{Dirac diffusion} \label{sec:dirac}

We consider the one-dimensional two-group radiation diffusion equation in a static medium with no coupling to the gas. The equations to solve are then
\begin{equation}
\begin{array}{@{}lcl@{}}
 \partial_t E_1 - \nabla \left( \frac{c}{3\rho \kappa_{\mathrm{R}1}} \nabla E_1 \right) = 0 \\
 \partial_t E_2 - \nabla \left( \frac{c}{3\rho \kappa_{\mathrm{R}2}} \nabla E_2 \right) = 0 ~.
\end{array}
\label{eq:dirac}
\end{equation}
For a constant $\rho\kappa_{\mathrm{R}}$ coefficient and a Dirac amplitude value of $E_0$ at $x_0$ as initial condition, the analytical solution $E_a$ in a $p$-dimensional space is
\begin{equation}
E_{a}(x,t)=\frac{E_0}{2^p (\pi \chi t)^{p/2}} e^{-\frac{(x-x_0)^2}{4\chi t}}
\end{equation}
where $\chi=c/(3\rho \kappa_{\mathrm{R}})$.

We choose a box of length $L=1$~cm, where $x_0=0.5$~cm. The gas has a uniform density $\rho=1~\text{g~cm}^{-3}$. The initial total radiative energy is set to $1~\text{erg~cm}^{-3}$ except in 
the center (inside the two central cells) where the peak value $E_{0}$ is set to $10^5~\text{erg~cm}^{-3}$. The two frequency groups' boundaries are (in Hz) [$10^{5}$, $10^{15}$] and [$10^{15}$,
$10^{19}$], chosen so that, in the central region, the radiative energy in the first group is about two orders of magnitude lower than in the second one. The Rosseland opacity in the first group 
is set to $\kappa_{\mathrm{R}1}=1~\text{cm}^2~\text{g}^{-1}$ and 10 times higher in the second group. The domain is initially divided into 32 cells (coarse grid level of 5) and 4 additional AMR 
levels are enabled (effective resolution of 512). The refinement criterion is based on the gradient of the total radiative energy. There is no flux-limiter, i.e. $\lambda_g=1/3$ for both 
groups, and a fixed timestep of $\Delta t = 2.5 \times 10^{-15}$~s is used for the coarse level (since we are using the adaptive time-stepping scheme of \citealp{Commercon2014}, the timestep is 
divided by two per AMR level increase).

\begin{figure}
\includegraphics[scale=0.38]{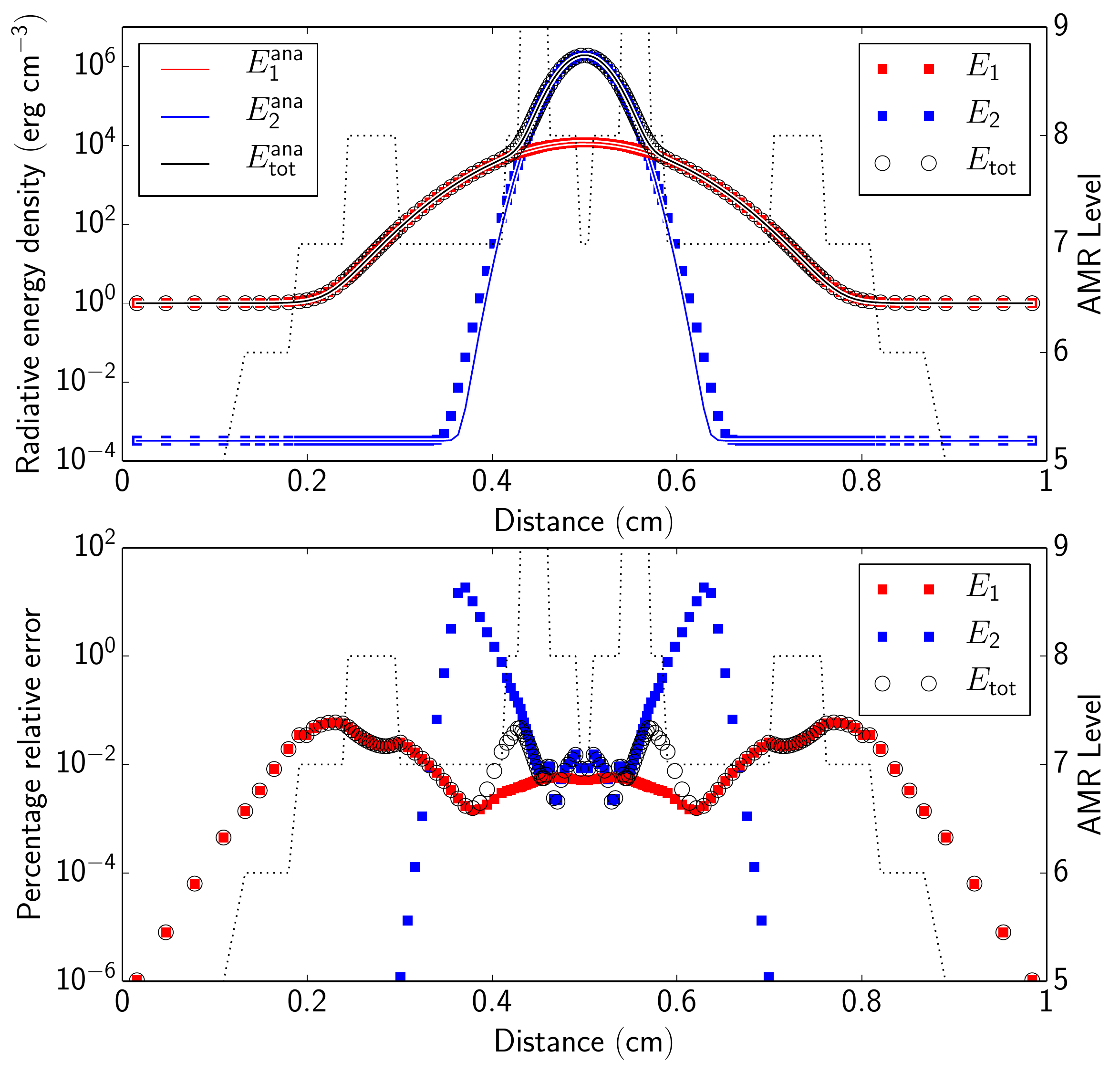}
\caption{Diffusion test: numerical (circles and squares) and analytical solutions (solid lines) at time $t=2 \times 10^{-13}$~s. Bottom panel: relative error. The dotted line corresponds to the 
AMR levels in the simulation.}
\label{fig:dirac}
\end{figure}

Figure~\ref{fig:dirac} (top) shows the radiative energy density profiles at time $t=2\times 10^{-13}$~s.
The two radiative energies are in good agreement with the analytical curves. The relative error (bottom panel) on the total radiative energy is always less than 10\%. The zones where the 
relative error on group 2 exceeds the 10\% mark correspond to regions where this group does not contribute to the total energy, making this error largely unimportant.

\subsection{Radiating plane} \label{sec:graziani}

\citet{Graziani2008} proposed an analytical multigroup test in spherical geometry. Let's consider a sphere of radius $R$, and temperature $T_{\mathrm{s}}$ which is surrounded by a cold medium at temperature $T_0<T_{\mathrm{s}}$, heat capacity $C_{\mathrm{v}}$, and spectral absorption coefficient $\rho\kappa_{\nu} = \sigma_{\nu}$. The test consists in computing the time-dependent spectrum at a given distance $r>R$ from the sphere center in the absence of any hydrodynamic motion. In the case where the heat capacity tends to infinity, the gas temperature is constant, and the spectrum is the superposition of the black-body spectrum of the cold medium and the spectrum of the radiation emitted by the hot sphere. The analytical solution for the spectral energy is
\begin{equation}
E_{\nu}=B(\nu,T_0)+\frac{R}{r}\left[B(\nu,T_{\mathrm{s}})-B(\nu,T_0)\right]F(\nu,r-R,t)
\end{equation}
where
\begin{eqnarray}
\nonumber F(\nu,d,t)=\frac{e^{-\sqrt{3} \sigma_{\nu} d}}{2} & \left[ \mathrm{erfc}\left(\frac12 \sqrt{\frac{3\sigma_{\nu}}{4ct}}d-\sqrt{ct\sigma_{\nu}}\right) \right.\\
&+ \left. \mathrm{erfc}\left(\frac12 \sqrt{\frac{3\sigma_{\nu}}{4ct}}d+\sqrt{ct\sigma_{\nu}}\right) \right] ~.
\end{eqnarray}

\begin{figure}
\centering
\includegraphics[scale=0.39]{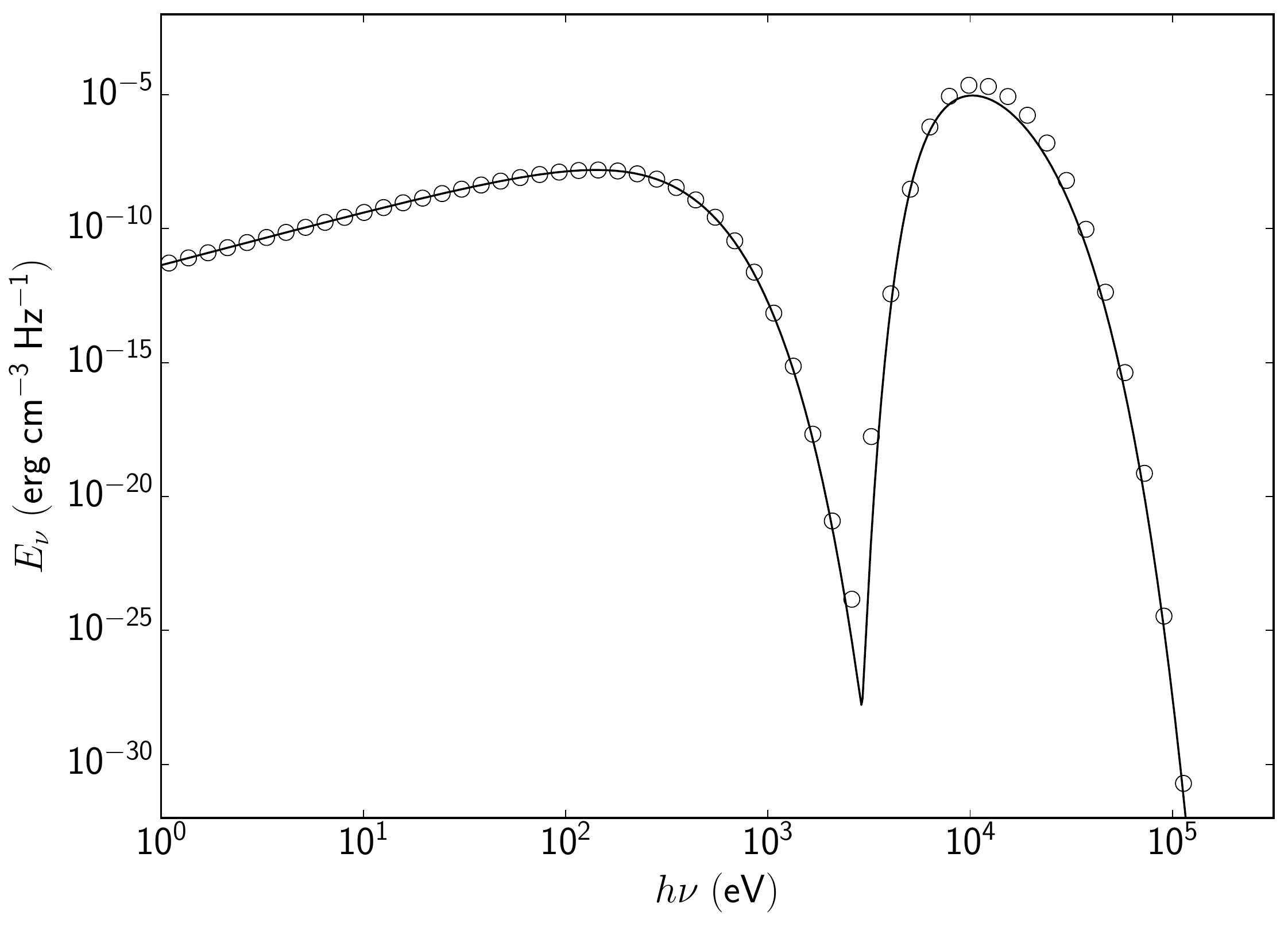}
\caption{Radiating plane test: numerical (circles) and analytical solutions at time $t=10^{-10}$~s.}
\label{fig:graziani}
\end{figure}

As our grid is Cartesian, we adapted this test to a slab geometry. Instead of a radiating sphere, we consider a radiating plane. The analytical solution can then be found making $r$ and $R$ 
tend to infinity, while keeping the distance $r-R$ constant \citep{Gentile2008}. We then simply have
\begin{equation}
E_{\nu}=B(\nu,T_0)+\left[B(\nu,T_{\mathrm{s}})-B(\nu,T_0)\right]F(\nu,r-R,t) ~.
\end{equation}
In our simulation, the temperature of the hot slab was set to $T_{\mathrm{s}}=1500$~eV and the medium was at $T_0=50$~eV.
The domain size is 0.1347368~cm with the hot slab located at the left boundary.
The domain was divided into 32 identical cells. We considered 60 groups logarithmically evenly spaced in the range [$0.5$~eV, $306$~keV], a fixed timestep of $\Delta t = 10^{-11}$ s was used, and no flux limiter (i.e. $\lambda_g=1/3$) was applied.
The gas absorption coefficient was set to $\sigma_g= 2 \times 10^{13} (\frac{h\nu_{g}}{1~\text{eV}})^{-3}$~cm$^{-1}$with $h\nu_{g}$ the energy in eV of the middle of each radiative group.
Figure~\ref{fig:graziani} shows the spectral radiative energies obtained compared to the analytical ones at a time $t=10^{-10}$~s, sampled at a distance $x = 0.04$~cm (this implies that
$r-R = 0.04$ cm in the analytical solution), which corresponds to the center of the tenth cell. The numerical results are in excellent agreement with the analytical solution.

\subsection{Non-equilibrium radiative transfer with picket fence model}\label{sec:suolson}

\citet{Su1999} developed analytical solutions for a 1D problem involving non-equilibrium radiative transfer 
with two radiative energy groups \citep[see also][]{Zhang2013}. Radiative energy is injected inside a small 
region of a uniform domain, diffuses and heats up the gas. The two groups have different opacities, so that 
the radiative energies propagate at different speeds through the medium. There are several assumptions in 
their analytical study. The heat capacity at constant volume is assumed to be $C_{\mathrm{v}} = 
d\epsilon/dT = \alpha T^{3}$ where $\alpha$ is a parameter. The group integrated Planck distribution is assumed to 
be
\begin{equation}\label{eq:PlanckSuOlson}
B_{g} = p_{g}\left(\frac{a_{\mathrm{R}}c}{4\pi}\right) T^{4} ~,
\end{equation}
where $a_{\mathrm{R}}$ is the radiation constant, and $p_{g}$ are parameters which verify the condition $\sum_{g} 
p_{g} = 1$. They then define the dimensionless coordinate $x = \bar{\sigma}z$ where $z$ is the coordinate
in physical units, and $\bar{\sigma} = \sum_{g} p_{g}\sigma_{g}$. The absorption coefficients $\sigma_{g}$
are independent of frequency, and scattering is ignored. The dimensionless time is
\begin{equation}\label{eq:tauDef}
\tau = \left(\frac{4a_{\mathrm{R}}c\bar{\sigma}}{\alpha}\right) ~,
\end{equation}
and the dimensionless radiative energy density and internal energy are
\begin{equation}\label{eq:UVDef}
U_{g} = \frac{E_{g}}{a_{\mathrm{R}}T_{0}^{4}} ~~~\text{and}~~~ V = \left(\frac{T}{T_{0}}\right)^{4} ~,
\end{equation}
respectively, where $T_{0}$ is a reference temperature.

The radiation source is applied for a finite period of time ($0 \le \tau < \tau_{0}$) inside the region
$|x| < x_{0}$, and gas dynamics are neglected. The equations solved are then
\begin{equation}
\begin{array}{@{}lcl@{}}
 \partial_t E_{\mathrm{T}} & = & \displaystyle \sum_{g=1}^{Ng}{\nabla \cdot \left(\frac{c\lambda_g}{\rho \kappa_{\mathrm{R}g}} \nabla E_g\right)}\\
\displaystyle \partial_t E_g - \nabla \cdot \left( \frac{c \lambda_g}{\rho \kappa_{\mathrm{R}g}} \nabla E_g \right) &=& \kappa_{\mathrm{P}g} \rho c \left(\Theta_g(T)-E_g+\Gamma_{g}\right) \\
\end{array}
\label{eq:suolson}
\end{equation}
where
\begin{equation}\label{eq:heating}
\Gamma_{g} = \left\{
\begin{array}{ll}
\displaystyle \frac{p_{g} \bar{\sigma} a_{\mathrm{R}} T_{0}^4}{\kappa_{\mathrm{P}g}\rho} & \text{if}~\tau < \tau_{0}~\text{and}~ |x| < x_{0} ~, \\
0 & \text{otherwise}~.
\end{array}
\right.
\end{equation}

\begin{figure}
\centering
\includegraphics[scale=0.36]{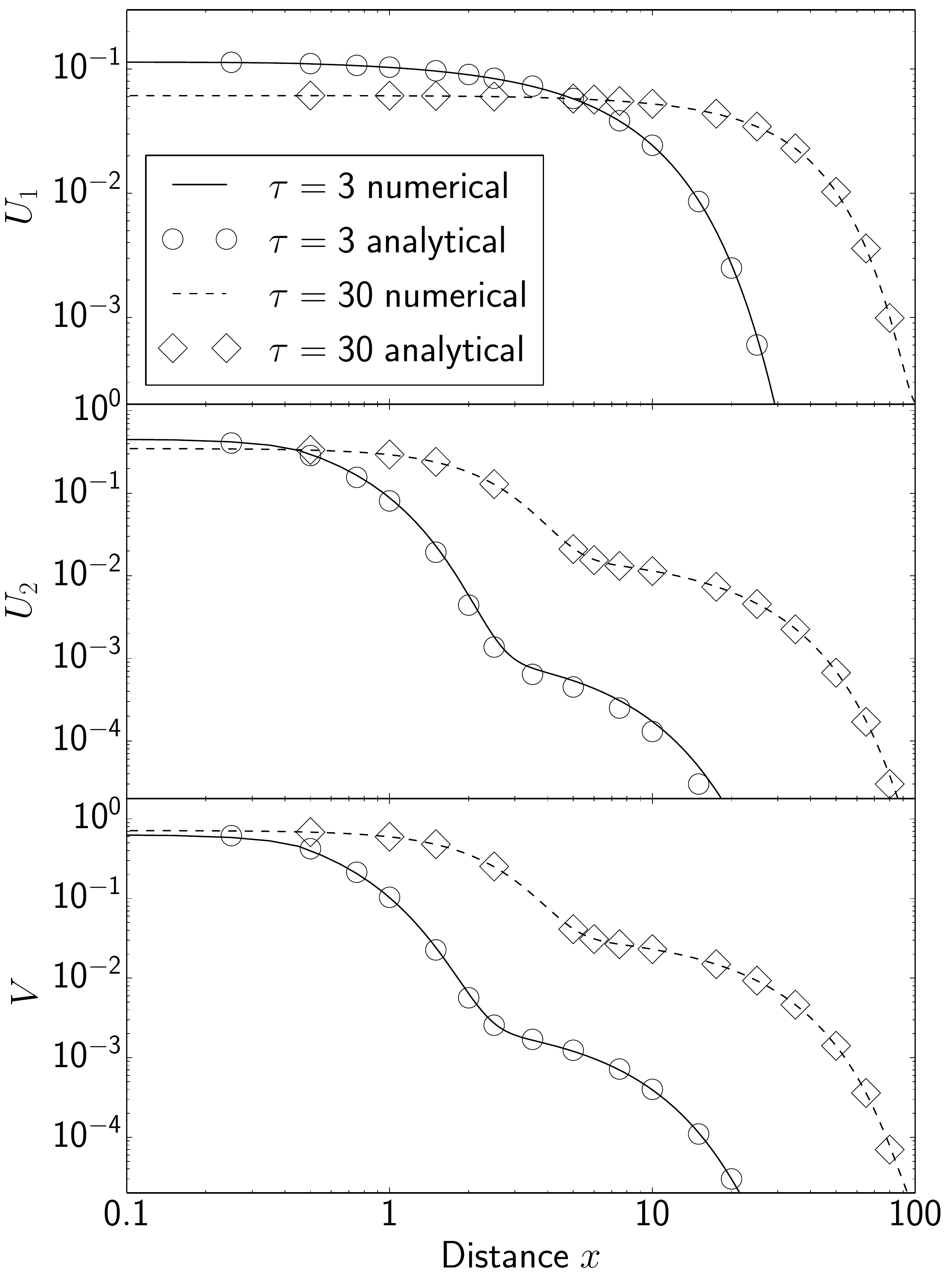}
\caption{Non-equilibrium radiative transfer test of \citet{Su1999}. Profiles from the numerical
simulation of $U_{1}$ (top), $U_{2}$ (middle) and $V$ (bottom) are shown for $\tau = 3$ (solid lines) and
$\tau = 30$ (dashed lines). The results are compared to the analytical solutions of \citet{Su1999} for
$\tau = 3$ (circles) and $\tau = 30$ (diamonds).}
\label{fig:suolson}
\end{figure}

Following \citet{Zhang2013}, we performed the case $\mathcal{C}$ of \citet{Su1999} and compared the results 
with their analytical solution for the radiation diffusion. In case $\mathcal{C}$ there are two radiation 
groups, with $p_{1} = p_{2} = 1/2$. The absorption coefficients are chosen as
$\sigma_{1} = 2/101~\text{cm}^{-1}$ and $\sigma_{2} = 200/101~\text{cm}^{-1}$, and the parameter $\alpha$
used to evaluate the heat capacity is $\alpha = 4a_{\mathrm{R}}$. The reference temperature is set to $T_{0} = 
10^{6}$ K. The radiation source parameters are $x_{0} = 1/2$ and $\tau_{0} = 10$. To avoid spurious 
boundary condition effects, we used a computational domain twice the size of \citet{Zhang2013}, spanning
$-102.4 < x < 102.4$, divided into 2048 identical cells. The left and right boundary conditions were both set to
periodic. The initial state of the physical variables
were $\rho = 1~\text{g~cm}^{-3}$ and $T = 1$ K, and matter and radiation were in equilibrium. A fixed timestep $\Delta \tau = 0.1$
was used, and no flux limiter (i.e. $\lambda_g=1/3$) was applied. The results are shown in 
Fig.~\ref{fig:suolson}, where an excellent agreement between the numerical and analytical solutions can be
seen.

\subsection{Radiative shocks with non-equilibrium diffusion} \label{sec:statshock}

\begin{figure*}
\centering
\includegraphics[scale=0.5]{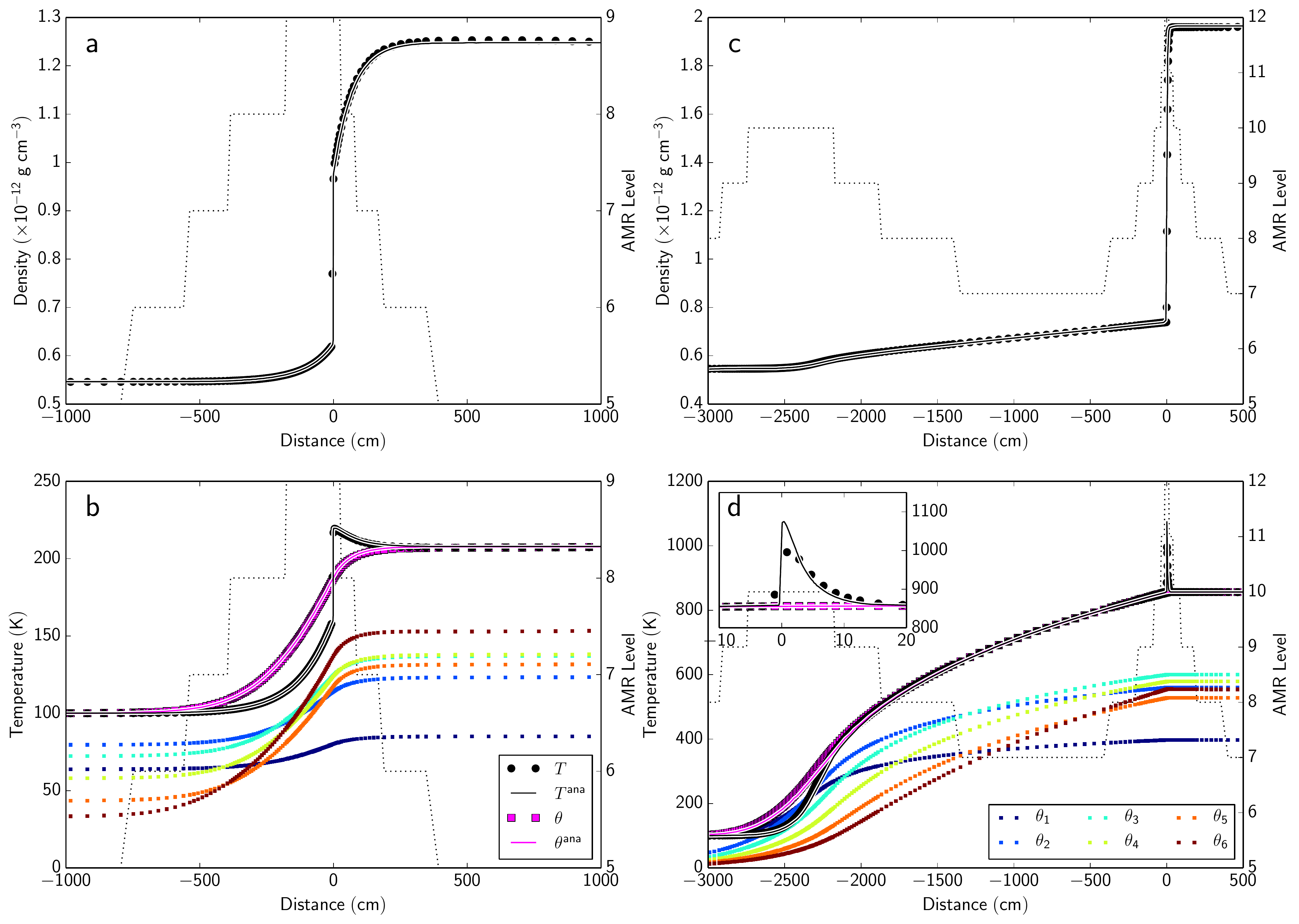}
\caption{Radiative shocks with non-equilibrium diffusion. The left column are the results for the $\mathcal{M} = 2$ case, while the right column is the $\mathcal{M} = 5$ setup. In all panels 
the numerical results are marked by symbols (circles and squares), the analytical solutions are represented by solid lines, and the mesh AMR level by dotted black lines. Panels (a) and (c) 
display the gas density as a function of distance. Panels (b) and (d) show the gas (black) and total radiative (magenta) temperatures. They also present the radiative temperatures inside the individual groups $\theta_{g}$ with more colours (see legend in panel d).}
\label{fig:radshocks}
\end{figure*}

The fourth test looks at the coupling between the fluid motion and the radiative transfer, solving the complete radiation-hydrodynamics equations (Eq.~\ref{eq:rhd}). \citet{Lowrie2008} developed
semi-analytic solutions to a 1D non-equilibrium diffusion problem involving radiative shocks of various Mach numbers $\mathcal{M}$. We simulated the $\mathcal{M} = 2$ and $\mathcal{M} = 5$ 
cases using 6 frequency groups. For both runs, the domain was split in two uniform regions where the hydrodynamic and radiation variables satisfy the Rankine-Hugoniot jump relations for a 
radiating fluid in an optically thick medium. The fluid variables inside the ghost cells at the left and right boundary conditions were kept as the initial pre- and post-shock state values 
throughout the simulation (imposed boundary condition). The gas has an ideal equation of state with a mean atomic weight $\mu=1$ and a specific heat ratio $\gamma=5/3$, and the Planck and 
Rosseland opacities are set to $\rho\kappa_{\mathrm{P}}=3.93 \times 10^{-5}~\text{cm}^{-1}$ and $\rho\kappa_{\mathrm{R}}=0.848902~\text{cm}^{-1}$, respectively, in all frequency groups. We used 
the HLL Riemann solver for the hydrodynamics with a CFL factor of 0.5 and no flux-limiter for the radiation solver (i.e. $\lambda_g=1/3$).

For the $\mathcal{M} = 2$ case, the initial conditions in the left (pre-shock) region are $\rho_{\mathrm{L}}=5.45887 \times 10^{-13}~\text{g~cm}^{-3}$, $u_{\mathrm{L}}=2.3547 \times 10^5~\text{cm~s}^{-1}$, 
$T_{\mathrm{L}}=100$~K and in the right (post-shock) region $\rho_{\mathrm{R}}=1.2479 \times 10^{-12}~\text{g~cm}^{-3}$, $u_{\mathrm{R}}=1.03 \times 10^5~\text{cm~s}^{-1}$, and $T_{\mathrm{R}}=207.757$~K. The domain ranges from -1000~cm 
to 1000~cm. Five frequency groups were evenly (linearly) distributed between 0 and $2\times 10^{13}$~Hz, and the sixth group ranged from $2\times 10^{13}$~Hz to infinity. The domain was 
initially divided in 32 cells and 4 additional AMR levels were enabled (effective resolution of 512). The refinement criterion was based on gas density and total radiative energy gradients. The 
density and temperature (gas and radiation) profiles are displayed in Figs~\ref{fig:radshocks}a and \ref{fig:radshocks}b. The thermodynamic quantities (gas density and temperature) are 
represented by the black symbols, while the radiation temperatures, defined by $\theta_{g} = (E_{g}/a_{\mathrm{R}})^{1/4}$, are marked by coloured squares. The numerical solutions show an 
excellent agreement with the semi-analytical solutions of \citet{Lowrie2008} (solid lines). The simulation data were shifted by 13.67 cm to place the density discontinuity at $x = 0$ for 
comparison with the analytical solutions; this corresponds to the shift the shock suffers as the radiative precursor develops, until the stationary state is reached.

In the $\mathcal{M} = 5$ case, the initial conditions are $\rho_{\mathrm{L}}=5.45887 \times 10^{-13}~\text{g~cm}^{-3}$, $u_{\mathrm{L}}=5.8868 \times 10^5~\text{cm~s}^{-1}$, $T_{\mathrm{L}}=100$~K, and
$\rho_{\mathrm{R}}=1.96405 \times 10^{-12}~\text{g~cm}^{-3}$, $u_{\mathrm{R}}=1.63 \times 10^5~\text{cm~s}^{-1}$, and $T_{\mathrm{R}}=855.72$~K. The domain ranges from -4000~cm to 4000~cm. Five frequency groups were 
evenly (linearly) distributed between 0 and $10^{14}$~Hz, and the sixth group ranged from $10^{14}$~Hz to infinity. The domain was initially divided in 32 cells and 
7 additional AMR levels were enabled (effective resolution of 4096). The refinement criterion was based on gas density, gas temperature, and total radiative energy gradients. The results in Figs~\ref{fig:radshocks}c and 
\ref{fig:radshocks}d show again an excellent agreement with the semi-analytical solutions. The simulation data were this time shifted by 193.5~cm to bring the density discontinuity to $x = 0$.

\section{Algorithm performance}\label{sec:scaling}

We present in this section some tests to assess the scaling performance of our algorithm.

\subsection{Strong and weak scaling}

\begin{figure}
\centering
\includegraphics[scale=0.37]{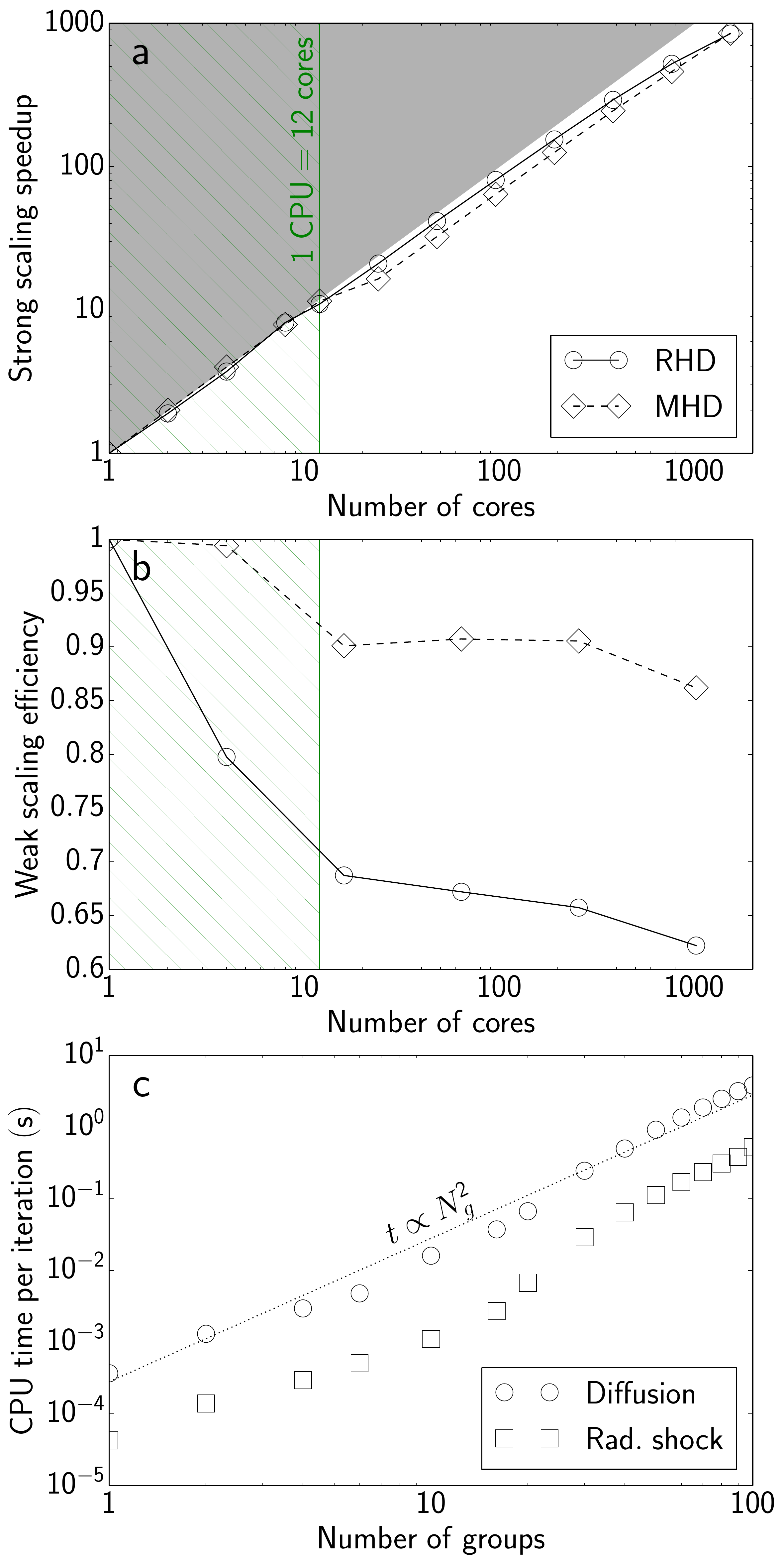}
\caption{(a) Strong scaling results for the multigroup diffusion test (circles) and the ideal MHD Orszag-Tang vortex calculation (diamonds). The ideal theoretical speedup is at the boundary
between the grey and white areas. The Occigen machine has 12-core processors; the 12-core region is marked by the green hatched area. (b) Weak scaling efficiency for the RHD and MHD runs.
(c) CPU time per BiCGSTAB iteration for a 1D diffusion problem (circles) and a 1D non-stationary radiative shock (squares) as a function of the number of frequency groups $N_{g}$ used. The 
curve for a time proportional to $N_{g}^{2}$ is shown with a dotted line.}
\label{fig:scaling}
\end{figure}

The strong and weak scaling runs were performed on the CINES Occigen\footnote{\url{https://www.cines.fr/calcul/materiels-et-logiciels/occigen/}} supercomputer, which uses 
Intel\textsuperscript{\textregistered} E5-2690 (2.60 GHz) processors. We compared the scaling of our radiative transfer scheme to the native MHD scheme in \texttt{RAMSES}. The setups used for 
the RHD and MHD runs were a 2D version of the Dirac diffusion test using 4 frequency groups and a 2D Orszag-Tang vortex simulation \citep{Orszag1979J}, respectively. In the 
RHD calculation, the fluid is static and only the radiation solver was called by \texttt{RAMSES}, bypassing the hydrodynamic Godunov solver. Both setups used a $2048^2$ mesh.

The strong scaling results are displayed in Fig.~\ref{fig:scaling}a. We can see that as we go beyond the 12-core limit of the Occigen processors, the speedups drop below the ideal curve (grey 
area), as communications begin to take longer to complete. Our radiative transfer method appears to perform slightly better than the native MHD scheme in \texttt{RAMSES}, as the coupling 
between hydrodynamics and radiation is ignored (the Godunov solver is not used in the RHD simulations). In the weak scaling 
RHD runs, when the size of the problem is doubled with the number of CPUs, it takes the implicit BiCGSTAB solver more iterations to converge, as there is a stronger propagation of round off 
errors originating from the calls to the \texttt{MPI\_ALLREDUCE} routine when more CPUs are used. To ensure a fair comparison, we forced all simulations to execute the same number of iterations 
(500) for every timestep, chosen as the maximum observed number of iterations in the 1024-core run. All simulations also performed the same number of timesteps (100) of a fixed length in time
($\Delta t = 10^{-17}$ s), and each CPU held a grid of $128^{2}$ cells. The weak MHD simulations were run for 300 timesteps, with each CPU processing a grid of $512^2$ cells\footnote{These 
resolutions and number of timesteps were chosen to that both RHD and MHD simulations run for approximately the same amount of time.}. The results in 
Fig.~\ref{fig:scaling}b show that the weak scaling performance of the RHD solver is below the native MHD solver. The iterative implicit solver suffers from heavy communication operations to 
compute residuals and scalar quantities which need to be performed at each iteration for each timestep, while the MHD solver only requires one communication operation per timestep. We believe 
that a weak scaling efficiency of 60\% remains however acceptable for our purposes.

\subsection{Group scaling}

Our final performance test was to assess the scaling of our multigroup algorithm for a given problem when the number of frequency groups $N_{g}$ is increased. The results of the simulation time 
divided by the total number of BiCGSTAB iterations for a 1D multigroup diffusion test performed on a single Intel\textsuperscript{\textregistered} Xeon\textsuperscript{\textregistered} E5620 
(2.40 GHz) CPU core on a local HP-Z800 workstation are shown in Fig.~\ref{fig:scaling}c (circles). The algorithm appears to scale with $N_{g}^{2}$, which is expected from the double sum over $N_{g}$ in the 
term on the third line of equation (\ref{eq:discrEr}). We carried out a second group scaling study, this time running the sub-critical radiative shock test from section~\ref{sec:statshock} 
(although with a lower resolution; only 3 levels of refinement were used\footnote{This explains a smaller time per iteration for the radiative shock simulations than for the diffusion runs.}), 
where the radiative transfer is fully coupled to the hydrodynamics. The results are displayed in Fig.~\ref{fig:scaling}c (squares), and the behaviour is very similar to the diffusion-only 
solver. The radiative transfer step completely dominates over the hydrodynamic step\footnote{The radiation solver takes up to 90\% of the computation time during one timestep.}, in terms of 
computational cost, and it is thus not surprising to see the same scaling for a RHD run than for a calculation which only calls the radiation solver.

\section{Application to star formation}  \label{sec:collapse}

In this final section, we apply the multigroup formalism to a simulation of the collapse of a gravitationally unstable dense cloud core, which eventually forms a protostar in its centre.
The collapsing material is initially optically thin and all the energy gained from compressional heating is transported away by the escaping radiation, which causes 
the cloud to collapse isothermally. As the optical depth of the cloud rises, the cooling is no longer effective and the system starts heating up, taking the core collapse 
through its adiabatic phase. A hydrostatic body \citep[also known as the first Larson core;][]{Larson1969} approximately 10~AU in size is formed and continues to accrete material from the 
surrounding envelope and accretion disk; this first core will eventually form a young star, after a second phase of collapse triggered by the dissociation of $\text{H}_{2}$ molecules 
\citep[see][for instance]{Masunaga2000}. In this preliminary astrophysical application, we shall however focus on the properties of the first Larson core for the sake of simplicity.

We adopt initial conditions similar to those in \citet{Commercon2010}, who follow \citet{Boss1979}. A magnetised uniform-density sphere of molecular gas, rotating about the $z$-axis with solid 
body rotation, is placed in a surrounding medium a hundred times less dense. The gas and radiation temperatures are 10 K everywhere. The prestellar core mass has a mass of 1 $\text{M}_{\odot}$, a radius
$R_{0} = 2500$ AU, and a ratio of rotational over gravitational energy of 0.03. To favor fragmentation, we use an $m = 2$ azimuthal density perturbation with an amplitude of 10\%. The magnetic 
field is initially parallel to the rotation axis. The strength of the magnetic field is expressed in terms of the mass-to-flux to critical mass-to-flux ratio $\mu = (M/\Phi)/(M/\Phi)_{c} = 5$ 
\citep{Mouschovias1976}. The field strength is invariant along the $z$ direction, and it is 100 times stronger in a cylinder of radius $R_{0}$ (with the dense core in its centre) than in the 
surrounding medium\footnote{This was chosen to try and reproduce the dragging-in of field lines that would have happened in the formation of the dense core \citep[see][for example]{Gillis1974}, 
while also retaining in the simplest manner the divergence-free condition for the MHD.}. We used a gas equation of state modeling a simple mixture of 73\% hydrogen and 27\% helium (in mass), which
takes into account the effects of rotational (for H$_{2}$ which is the dominant form of hydrogen for temperatures below 2000 K) and vibrational degrees of freedom. The frequency-dependent dust 
and gas opacities were taken from \citet{Vaytet2013a}, assuming a 1\% dust content. We used the ideal MHD solver of \texttt{RAMSES}, and the grid refinement criterion was based on the Jeans
mass, ensuring the Jeans length was always sampled by a minimum of 12 cells. The coarse grid had a resolution of $32^{3}$, and 11 levels of AMR were enabled, resulting in a maximum resolution of 0.15 AU at the finest level. The Minerbo flux limiter \citep{Minerbo1978} was used for these simulations.

\begin{figure}
\centering
\includegraphics[scale=0.315]{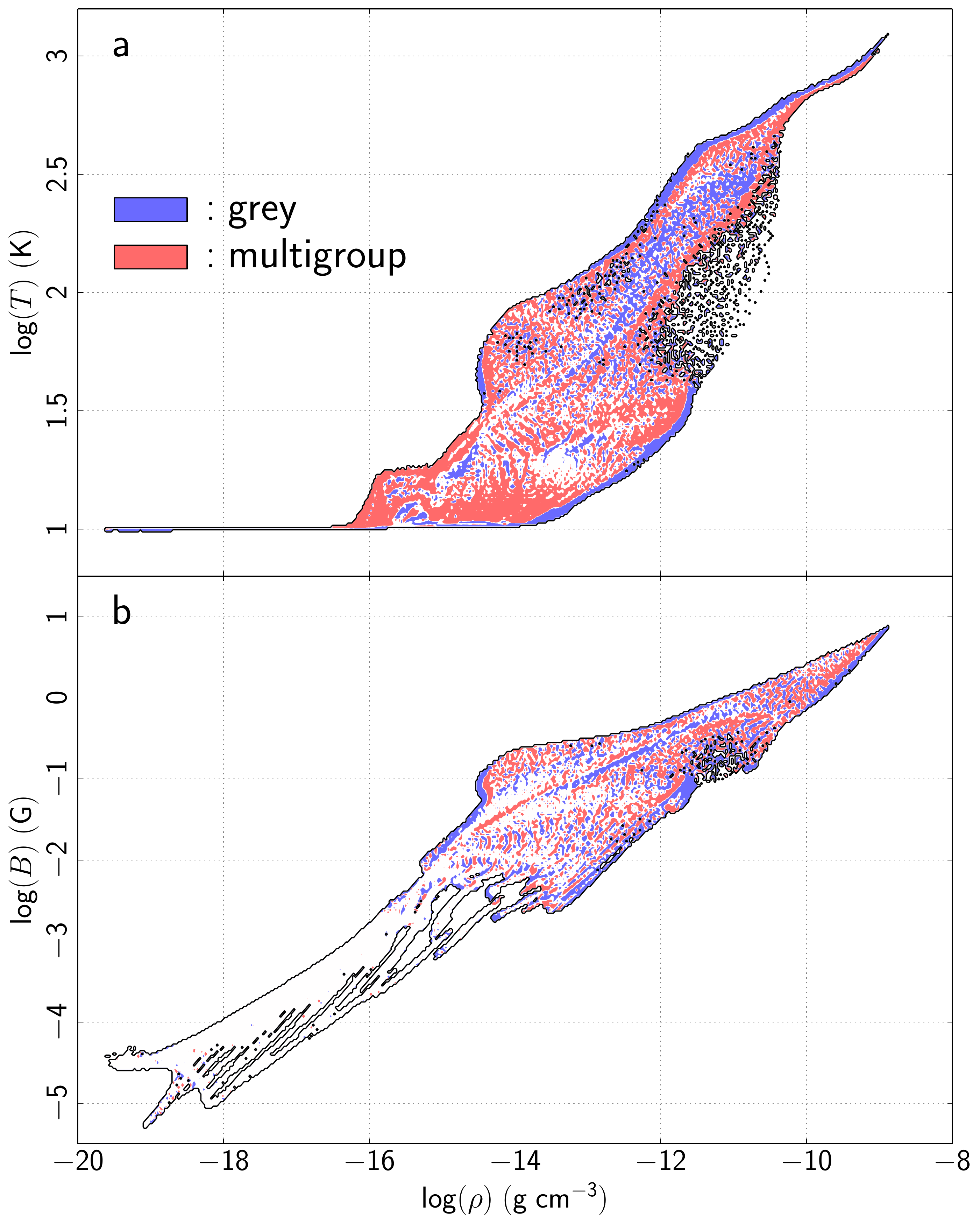}
\caption{(a) Temperature as a function of density in the grey and multigroup simulations at a time $t$ = 24,265 yr. The blue colour represents regions where the grey simulation either 
dominates (in terms of mass contained within the figure pixels) over the multigroup simulation, or where there is no multigroup data. Likewise, the red codes for the regions of the diagram 
where the multigroup run prevails. The white areas are where both simulations yield identical results. The black contour line delineates the region where data are present. (b) Same as for (a) 
but showing the strength of the magnetic field as a function of density.}
\label{fig:trho_brho}
\end{figure}

\begin{figure*}
\centering
\includegraphics[scale=0.50]{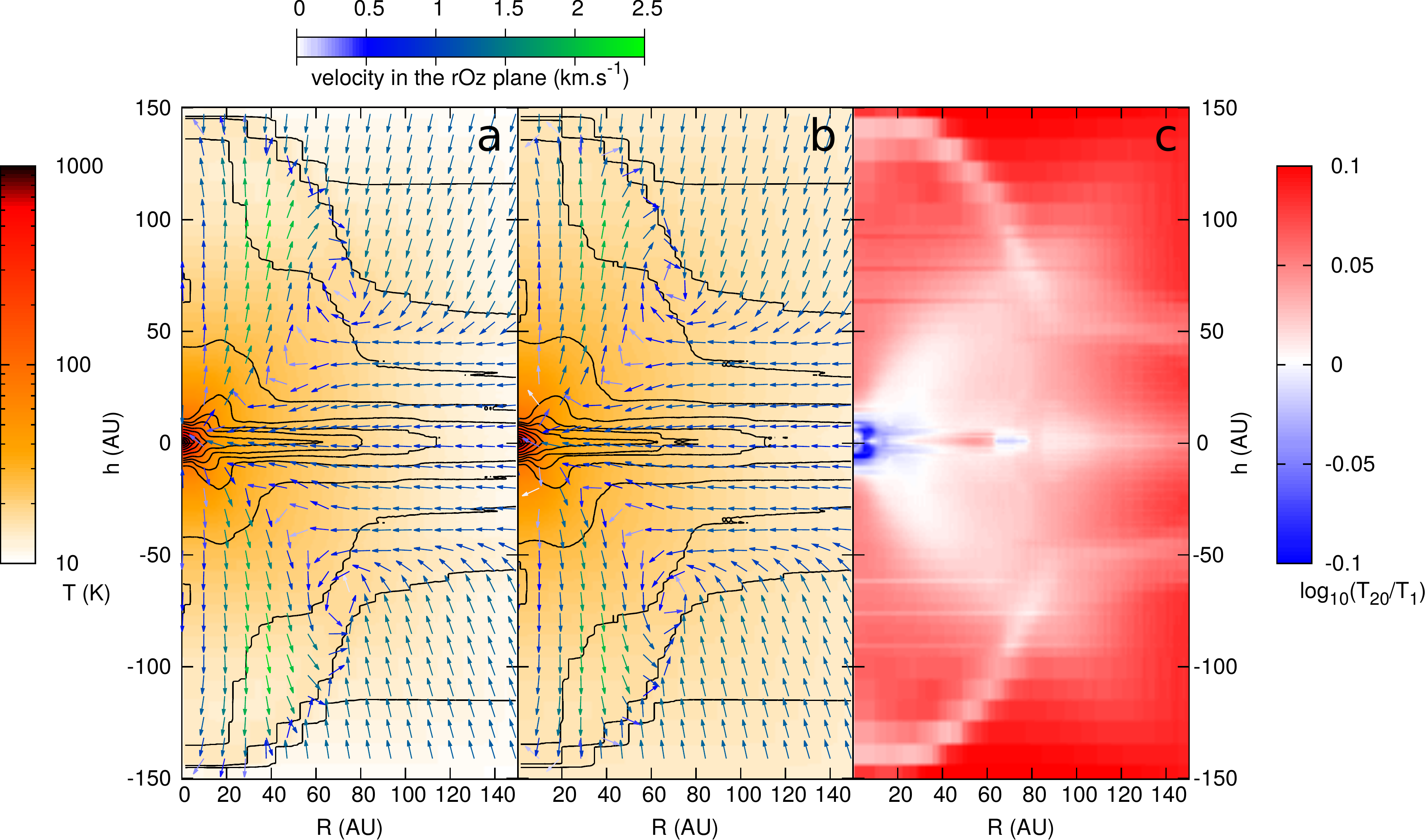}
\caption{(a) Temperature map (colours) at a time $t$ = 24,282 yr in a grey simulation of the collapse of a $1~\text{M}_{\odot}$ cloud. The data are presented as a function of radius and height 
over the mid-plane; the data have been averaged around the azimuthal direction. The black lines are logarithmically spaced density contours, in the range
$10^{-16} < \rho < 10^{-9}~\text{g~cm}^{-3}$ (two contours per order of magnitude). The arrows represent the velocity vector field in the $r$-$z$ plane. (b) Same as for (a) but using 20 
frequency groups. (c) Map of the logarithm of the ratio of the multigroup temperature over the grey temperature. The red colour indicates where the gas temperature in the multigroup simulation 
is higher than for the grey run, and vice-versa for the blue colour.}
\label{fig:Tmaps}
\end{figure*}

We performed two simulations; one under the grey approximation and a second using 20 frequency groups. The first and last groups spanned the frequency ranges (in Hz) $[0 \rightarrow 5\times10^{10}]$ and
$[1.3\times10^{14} \rightarrow \infty]$, respectively. The remaining 18 groups were evenly (logarithmically) distributed between $5\times10^{10}$ and $1.3\times10^{14}$ Hz.
The results are shown in Figs~\ref{fig:trho_brho} and \ref{fig:Tmaps}. The gas 
temperature as a function of density for all the cells in the computational domain is shown in Fig.~\ref{fig:trho_brho}a, where the blue colour represents 
regions where the grey simulation dominates (in terms of mass contained within the plot pixels) over the multigroup simulation, and the red shows where multigroup data prevail. While the two
simulations show similar results overall, there are several differences we wish to point out. For relatively low densities in the range
$5 \times 10^{-17} < \rho < 3 \times 10^{-15}~\text{g~cm}^{-3}$, 
the multigroup run is hotter than the grey simulation. This is also visible in the temperature maps of Fig.~\ref{fig:Tmaps}, where the gas is hotter in the 20-group run for
$r > 20$ AU, this being most obvious in panel (c). It appears that the radiation transport from the central core to the surrounding envelope is more efficient in the multigroup case. More 
energy has left the core, rending it colder than in the grey case, while more energy has been deposited in the thus warmer outer envelope. Higher temperatures are also observed in the bipolar 
outflow, along the vertical axis, relatively close to the first core. \citet{Kuiper2011} found that using a frequency-dependent scheme in simulations of high-mass stars could enhance radiation 
pressure in the polar direction compared to the grey simulations of \citet{Krumholz2009}, producing much stabler outflows. This could be similar to what we are observing here, although this 
requires further study.

Conversely, the strength of the magnetic field does not change significantly when using a multigroup model. In fact, for densities below $10^{-15}~\text{g~cm}^{-3}$, the two runs are virtually
identical (see Fig.~\ref{fig:trho_brho}b). In the ideal MHD limit, the magnetic field is not directly related to the thermal properties of the gas, and it is therefore not surprising that the
multigroup formalism has little impact. However, we plan in the future to study star formation using a non-ideal description of MHD \citep{Masson2012}, where magnetic and thermal interactions
are twofold. First, magnetic diffusion contributes additional gas heating from ion-neutral frictions and Joule heating. Second, the chemical properties of the gas, which strongly depend on 
temperature, also impact the magnetic resistivities, which govern the diffusion processes. We will investigate this in detail in a forthcoming study.

\begin{figure}
\centering
\includegraphics[scale=0.50]{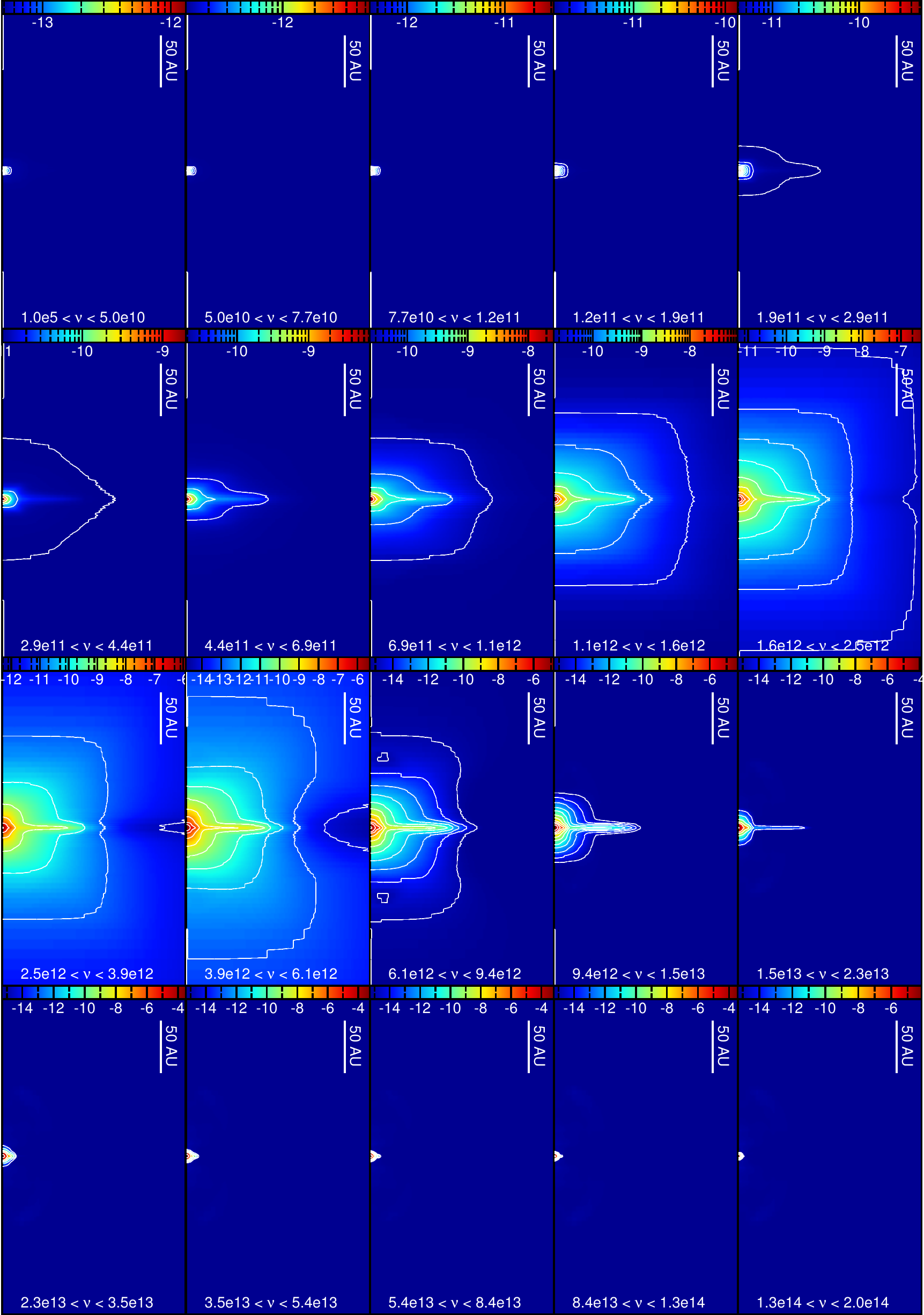}
\caption{Radiative energy density map in each frequency group. The white contours have 10 levels logarithmically spaced between the minimum and maximum values of each map. The frequencies of
each group are indicated at the bottom of each map. The maps represent the same region as in Fig.~\ref{fig:Tmaps}.}
\label{fig:Cmaps}
\end{figure}

\begin{figure}
\centering
\includegraphics[scale=0.30]{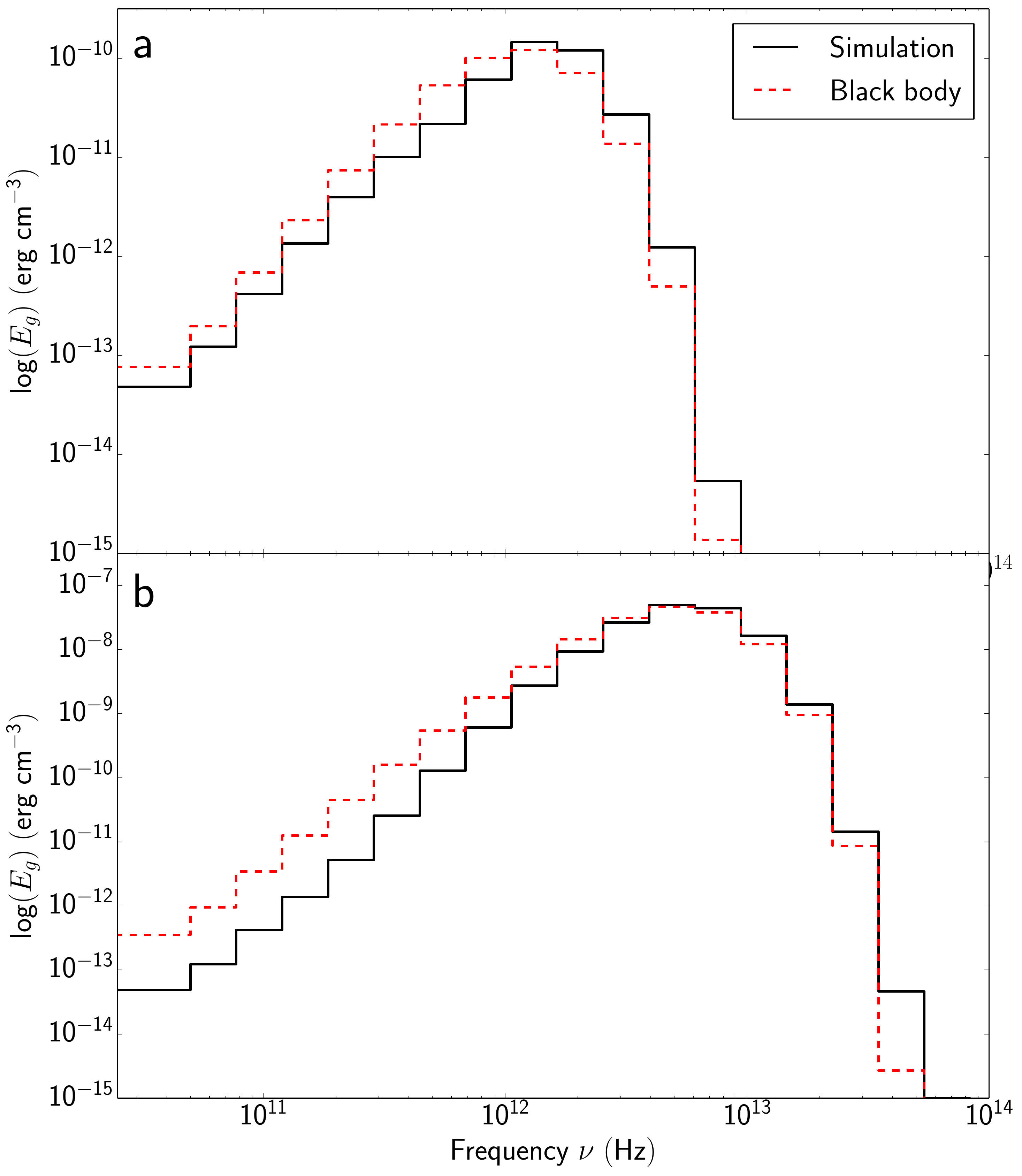}
\caption{Spectral energy distributions extracted in a cell located 2000~AU (a) and 20~AU (b) from the protostar. The black solid lines represent the energy inside the 20 frequency groups. They are compared to
a black body distribution (dashed red) which would have the same total energy.}
\label{fig:seds}
\end{figure}

The use of multigroup RHD may not yield significantly different results in the early stages of molecular cloud collapse, but it does provide a wealth of physical information in the system.
Channel maps such as the ones presented in Fig.~\ref{fig:Cmaps} or spectral energy distributions (SEDs; Fig.~\ref{fig:seds}) are directly available from the simulation data, without requiring
any post-processing software. The channel maps show a peak intensity around a wavelength of 100 $\mu$m, close to the synthetic observations of \citet{Commercon2012}. The SEDs interestingly show 
departures from a black body spectrum, this being most obvious at the low-frequency end, close to the protostar (20 AU;
Fig.~\ref{fig:seds}b). We do not wish here to carry out a detailed study on the effects of multi-frequency radiative transfer on the structures of protostars, we simply wish to illustrate the
power of the method. We leave the detailed work on collapsing objects for a future paper, as this is first and foremost a methodology focused article.

\section{Conclusions and future work}\label{sec:conclusions}

We have implemented a method for multigroup flux-limited diffusion in the \texttt{RAMSES} AMR code for astrophysical fluid dynamics.
The method is based on the time-implicit grey FLD solver of \citet{Commercon2011b}, and uses the adaptive time-stepping (in which each level is able to evolve with its own timestep using a subcycling procedure) strategy of \citet{Commercon2014}.
The multigroup method allows the discretisation of the frequency domain to any desired resolution, enabling us to take into account the frequency dependence of emission and absorption coefficients.
The radiative energy density in the frequency groups are all coupled together through the matter temperature and terms reproducing Doppler shift effects when velocity gradients are present in the fluid.
A consequence of this coupling is the apparition of non-symmetric terms in the matrix we have to invert in our implicit time-stepping procedure. We therefore had to abandon the original conjugate gradient algorithm of \citet{Commercon2011b} for a bi-conjugate gradient
iterative solver. A more evolved BiCGSTAB solver was preferred for its greater stability compared to a raw bi-conjugate algorithm.

The method was fully tested against standard radiation diffusion, frequency-dependent, and full radiation hydrodynamics tests. It performed extremely well in all of these tests, and its scaling performance was also found to be very satisfactory.

The multigroup formalism was finally applied to a simulation of the gravitational collapse of a dense molecular cloud core in the context of star formation. The method has revealed differences between grey and frequency-dependent simulations, but more importantly uncovered departures from
a black-body radiation distribution. We also illustrated the wealth of information the method brings to astrophysical studies, with the ability to directly produce channel maps and SEDs.
We will carry out a much more thorough study of the effects of multigroup radiative transfer on the structures of protostars and proto-planetary discs, as well as their observable quantities, as part of a much wider parameter space study in a forthcoming paper.

\begin{acknowledgements}
This work was granted access to the CINES HPC resources Jade and Occigen under the allocations x2014-047247 and x2015-047247 made by DARI/GENCI. The research leading to these results has also 
received funding from the European Research Council under the European Community's Seventh Framework Programme (FP7/2007-2013 Grant Agreement no. 247060). BC gratefully acknowledges support
from the French ANR Retour Postdoc program (ANR-11-PDOC-0031). We finally acknowledge financial support from the ``Programme National de Physique Stellaire'' (PNPS) of CNRS/INSU, France.
\end{acknowledgements}

\bibliographystyle{aa}

\end{document}